\documentclass{mrf}

\usepackage{graphics} 
\usepackage{amsmath}
\usepackage{subfig}
\usepackage{epstopdf}

\begin{document}

\supertitle{research paper}

\title[Radar for projectile impact]{Radar for projectile impact on granular media}

\author[Radar for projectile impact] {Felix Rech$^{1}$ and Kai Huang $^{1,2}$}

\address{\add{1}{Experimentalphysik V, Universit\"at Bayreuth, 95440 Bayreuth, Germany}
\add{2}{Division of Natural and Applied Sciences, Duke Kunshan University, No. 8 Duke Avenue, Kunshan, Jiangsu, China 215316}}

\corres{\name{Kai Huang}
\email{kh380@duke.edu}}

\begin{abstract}
From the prevention of natural disasters such as landslide and avalanches, to the enhancement of energy efficiencies in chemical and civil engineering industries, understanding the collective dynamics of granular materials is a fundamental question that are closely related to our daily lives. Using a recently developed multi-static radar system operating at $10$\,GHz (X-band), we explore the possibility of tracking a projectile moving inside a granular medium, focusing on possible sources of uncertainties in the detection and reconstruction processes. On the one hand, particle tracking with continuous wave radar provides an extremely high temporal resolution. On the other hand, there are still challenges in obtaining tracer trajectories accurately. We show that some of the challenges can be resolved through a correction of the IQ mismatch in the raw signals obtained. Consequently, the tracer trajectories can be obtained with sub-millimeter spatial resolution. Such an advance can not only shed light on radar particle tracking, but also on a wide range of scenarios where issues relevant to IQ mismatch arise.
\end{abstract}

\keywords{ Authors should not add keywords, as these will be chosen during the submission process (see http://journals.cambridge.org/\-data/\-relatedlink/\-MRF\_topics.pdf for the full list)}

\maketitle

\section{Introduction}

As large agglomerations of macroscopic particles, granular materials are ubiquitous in nature, industry and our daily lives \cite{Duran2000, pg13}. Despite of its importance, a fundamental understanding of granular dynamics from the perspective of transitions from a solid- to a liquid-like state (e.g., when and where does an avalanche start) is still far from complete. One of the main challenges in deciphering the dynamics of granular materials arises from the fact that most granular particles are opaque. In the past decades, there have been substantial progresses in imaging granular particles \cite{Amon2017}. Optical means for imaging particles in three dimensions (3D), such as refractive index matching \cite{Dijksman2017}, are limited to certain combinations of particles and interstitial liquids. X-ray tomography \cite{Athanassiadis2014} and Magnetic Resonance Imaging \cite{Stannarius2017} are also frequently used to identify the internal structures of granular materials. However, the limited time resolution of scanning technique as well as the huge amount of data to be processed hinder the investigation of granular dynamics, which is better resolved with sufficiently high temporal resolution. Note that the mobility of single particles can influence dramatically the collective behavior, owing to its discrete nature as well as heterogeneous distributions of force chains inside \cite{Bi2011}. Therefore, it is desirable to have a technique capable of tracking granular particles with high temporal resolution. 

Since the beginning of last century, radar technology has been continuously developed and benefiting us in many different ways: From large scale surveillance radar systems that are crucial for aircraft safety and space exploration \cite{Skolnik2001}, to small scale systems for monitoring insects \cite{Oneal2004}. Considering the capability of radar tracking technique, it is intuitive to ask: How small can an object be accurately tracked by a radar system? Can it be as small as a tracer particle with a size comparable to a grain of sand? Recently, we introduced a small scale continuous-wave (CW) radar system working at X-band to track a spherical object with a size down to $5$\,mm \cite{Ott2017}. In comparison to other techniques, the continuous trajectory of a tracer particle obtained by the radar system helps in deciphering granular dynamics greatly, owing to the high temporal resolution. Here, advantages and disadvantages of this technique, particularly how to handle the possible sources of uncertainty will be discussed in details.

\section{Particle tracking set-up and reconstruction algorithm}
\begin{figure}
\centering
\includegraphics[width=0.85\columnwidth]{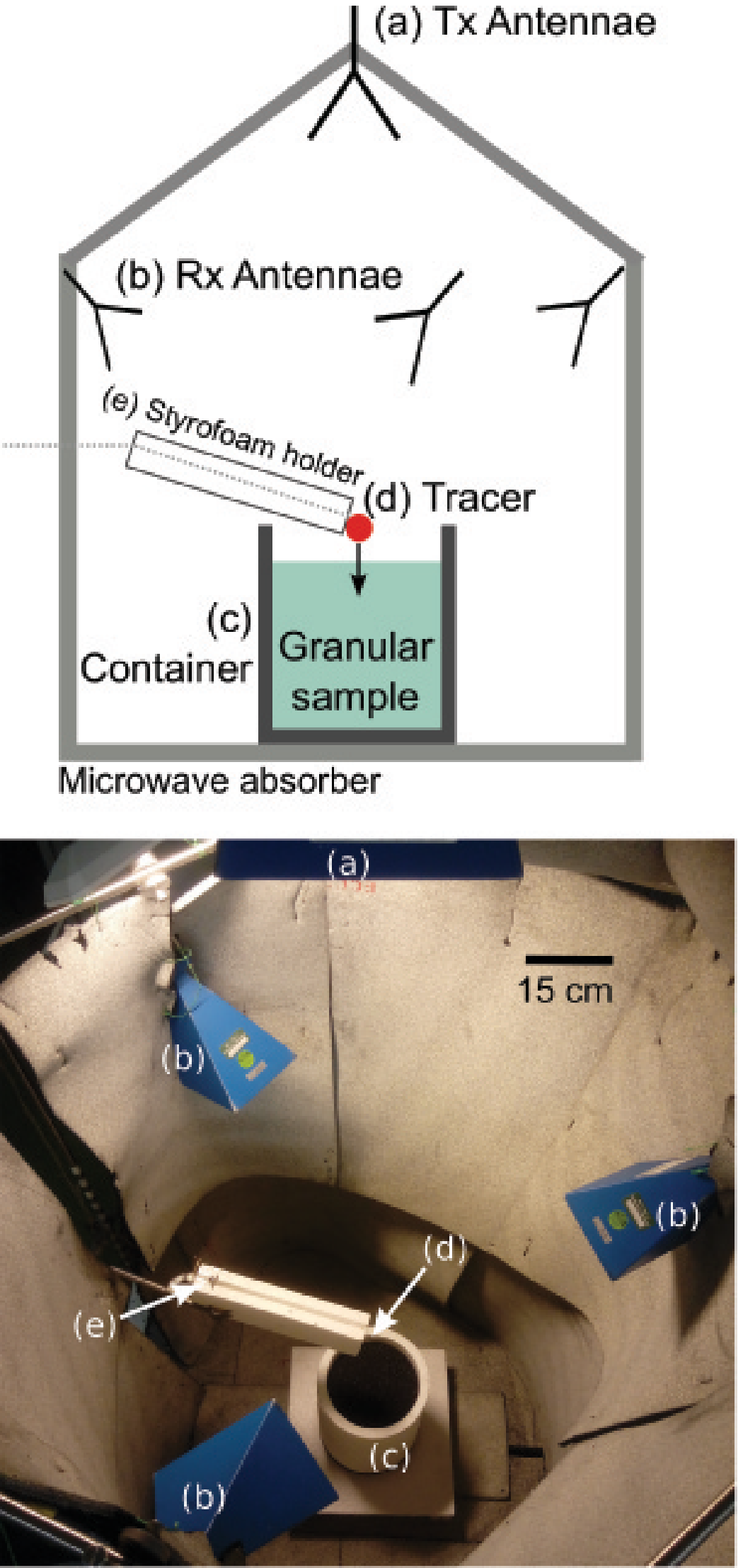}
\caption{Schematics (upper panel) and a top-view picture of the particle tracking system, showing the configuration of the (a) transmission (Tx.) and (b) receiving (Rx.) antennae, (c) granular sample, (d) tracer as well as the (e) tracer holding and releasing device.}
\label{fig:setup}
\end{figure}

Figure~\ref{fig:setup} shows the experimental set-up that utilizes the radar system to track a free falling projectile into a granular bed. The multi-static radar system operates at 10\,GHz (X-band) with one transmission (Tx.) antenna pointing in the direction of gravity (defined as $-z$ direction) and three receiving (Rx.) antennae mounted symmetrically around the $z$ axis. Polarized electromagnetic (EM) waves, after being scattered by the tracer particle, are captured by the Rx. antennae. 

A metallic sphere with a diameter $d=$10\,mm is used as the tracer. The cylindrical container has an inner diameter of $15$\,cm and a height of $20$\,cm. To avoid unnecessary multiple scattering, the sample holder is constructed with mostly foamy materials as possible. The granular sample (see Fig.\,\ref{fig:granulates} and relevant text for more details) is filled in the container till few centimeters below the rim and gently tapped until the initial packing density of $60.2$\% is achieved. The tracer is initially held by a thin thread wrapped around and released by gently pulling the thread such that the initial velocity of the falling sphere is close to $0$. This design enables a defined and reproducible initial condition for a comparison among various experimental runs. The raw IQ signals from the AD converter (NI DAQPAD-6015) are recorded and further processed with a Matlab program to obtain the reconstructed trajectories.

IQ-Mixers play an essential role in accurate ranging a target, as it measures the phase shift (i.e., time delay) of the received signal with respect to the emitted one. Suppose the latter can be described as $a\cos(2\pi f_{\rm 0}t)$ and the former as $b\cos(2\pi ft + \theta)$, where $a$ and $b$ are the magnitudes of the corresponding signals, $f_{\rm 0}$ and $f$ are the transmitted and received signal frequencies, the output signals of an IQ-Mixer is

\begin{eqnarray}
I=\frac{ab}{2}\cos[2\pi(f_{\rm 0}-f)t - \theta], \nonumber \\
Q=\frac{ab}{2}\sin[2\pi(f_{\rm 0}-f)t - \theta]. \label{eq:iq2}
\end{eqnarray}

\noindent Subsequently, the relative movement of the tracer is obtained from the phase shift of $I+Qi$ in a complex plane, where I (in-phase) and Q (quadrature) correspond to the two outputs of a IQ mixer. With the help of an IQ mixer, the change of the absolute traveling distance for the $i$th antenna $L_i = l_0 + l_i$ can be obtained, where $l_0$ and $l_i$ are the distance between Tx. antenna and the target and that from the target to the Rx. antenna, respectively. If $L_i$ varies by a distance of one wavelength, the vector $I+Qi$ rotates $2\pi$. As IQ mixers provide analogue signals representing the mobility of the tracer, the time resolution of the radar system is only limited by the analogue-digital (AD) converter. This is an advantage of using continuous wave radar systems for particle tracking.

Although distance measurements rely only on the phase information, the sensitivity and accuracy of the system depend on IQ signal strength. In order to have a sufficient signal to noise ratio, the directions of the horn antennae (Dorado GH-90-20) are adjusted with the help of a laser alignment and range meter (Umarex GmbH, Laserliner) to face the target area. The three Rx. antennae are arranged symmetrically about the axis starting from the Tx. antenna pointing along the direction of gravity, so that the received signals have similar signal-to-noise ratios. According to the specification of the antenna, the main lobe of its radiation pattern has an opening angle of $\sim15$ degrees. Thus, we estimate the field of `view' of the radar system has a volume of about $30 {\rm cm} \times 30 {\rm cm} \times 30 {\rm cm}$, taking into account the average working distance of the antennae. Nevertheless, as will be shown below, the field of view can be extended after a proper correction of the IQ mismatch. The distance between each antenna and the center of the coordinate system is also measured by the laser meter during the adjustment process. The polarization of the antennae are adjusted to maximize the raw I and Q signals. The whole system is covered with microwave absorbers (Eccosorb AN-73) to reduce clutter and unwanted noises from the surrounding.

From the measured distances $\vec{L} \equiv (L_1, L_2, L_3)$, the reconstructed trajectory can be obtained with a coordinate transformation

\begin{equation}
\label{eq:reconst}
\left(
\begin{array}{c}
x \\
y \\
z
\end{array}
\right)=\frac{\vec{r}-\vec{L}}{\vec{T}},
\end{equation}

\noindent where the vector $\vec{r}$ is chosen to be $0$ as it contributes only to a constant offset to the reconstructed trajectory, the transformation matrix reads

\begin{equation}
\label{eq:trmat}
\vec{T} \equiv \left(
\begin{array}{ccc}
\sin{\theta_{\rm 1}}\cos{\phi_{\rm 1}} & \sin{\theta_{\rm 1}}\sin{\phi_{\rm 1}} & 1+\cos{\theta_{\rm 1}} \\
\sin{\theta_{\rm 2}}\cos{\phi_{\rm 2}} & \sin{\theta_{\rm 2}}\sin{\phi_{\rm 2}} & 1+\cos{\theta_{\rm 2}} \\
\sin{\theta_{\rm 3}}\cos{\phi_{\rm 3}} & \sin{\theta_{\rm 3}}\sin{\phi_{\rm 3}} & 1+\cos{\theta_{\rm 3}} 
\end{array}\right)
\end{equation} 

\noindent with $\theta_i$ and $\phi_i$ the tilt and azimuth angles of the $i$th antenna, respectively. The transformation matrix is determined from a calibration process using the same tracer particle moving in a given circular trajectory in the horizontal plane. More descriptions of the calibration and coordinate transformation processes can be found in \cite{Ott2017}.

\section{How to enhance spatial resolution?}

Although extremely high temporal resolution can be achieved with the CW radar system, the spatial resolution relies strongly on the adjustment of the system, as well as the calibration and reconstruction algorithms. There are three main sources of uncertainty in the calibration and reconstruction process: (i) Fitting error arising from the calibration process; (ii) Reflection and multiple scattering from surroundings; (iii) Mismatch between signals obtained by I and Q channels. In the following part of this session, details on how to handle various sources of uncertainty, particularly for the case of IQ mismatch, will be discussed.

\begin{figure}
\centering
\includegraphics[width=0.95\columnwidth]{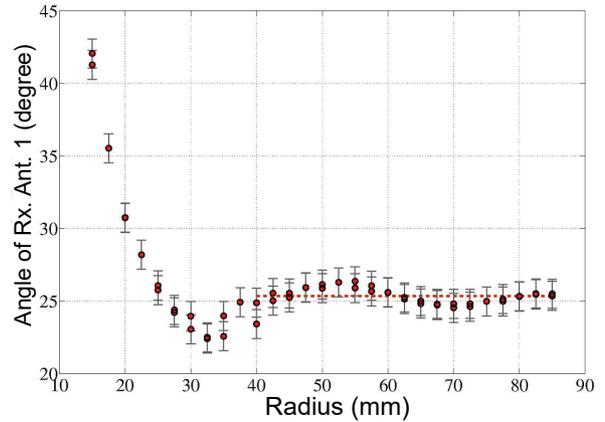}
\caption{The elevation angle of antenna 1 obtained from the calibration process as a function of the radius of circular trajectory. The red dashed line marks the averaged angle from Radius $40$ mm to $85$ mm. The error bars correspond to the uncertainty from the fitting algorithm.}
\label{fig:cali}
\end{figure}

First, the antennae parameters determined from the calibration process are essential for the accurate reconstruction of the tracer trajectories. Following the algorithm described above, circular motion of the tracer particle in the calibration process leads to harmonic oscillations of $L_{\rm i}$. Based on a first order approximation \cite{Ott2017}, the tilt and azimuth angle $\theta_i$ and $\phi_i$ of the Rx. antennae can be determined from fitting. Note that although the antenna parameters can be directly measured by laser assisted alignment tools, the outcome can not only serve as a rough initial guess, since an accurate determination of the exact location where the electromagnetic waves are emitted is nontrivial. Thus, we need to apply the aforementioned fitting algorithm to a reference trajectory. We choose a circular trajectory in the horizontal plane (defined as $x-y$ plane) for the following reasons: (a) The center of the three dimensional Cartesian system can be defined as the center of rotation with the rotating axis pointing toward the $+z$ direction, along which the Tx. antenna is aligned. In another word, the coordinate system is defined by the calibration circle. (b) A circular trajectory with different radius $R$ and rotation frequency $f$ is implemented with a Styrofoam tracer holder attached to a stepper motor. Styrofoam, which shares similar material properties as granular sample, is chosen as they are transparent to EM waves and rigid enough to support the tracers. Here the tracer is directly embedded into the rotating arm. Concerning the accuracy of determining antennae parameters, the circular trajectory has to be well constructed. Due to the low rigidity of Styrofoam, sources of errors such as eccentricity of the trajectory due to the coupling with the motor shaft or the relative motion of the tracer with respect to the Styrofoam arm may occur and lead to higher relative error in the radius of the reference circle, which we consider to be one source of error. As shown in Fig.\,\ref{fig:cali}, the fitted outcome of tiling angle of Rx. antenna 1 deviates systematically away from the expected value as $R$ decreases. Note that the calibration outcome should not depend on the radius and angular frequency of the circular motion. The above comparison suggests that the radius has to be larger than $\approx 40$\,mm for a reliable determination of the antennae parameters for the current configuration.

Second, there always exist scattered signals from the surrounding environment, thus it is necessary to have EM absorbing materials to enhance the signal-to-noise ratio. In addition, microwave absorbers also play an important role in isolating the system from the surrounding environment as the Rx. antennae may response to people walking around. In practice, the most efficient way of signal enhancement is to record a background signal at the same configuration of the experimental setup and subtract the background signal for the I and Q signals of all three channels. As shown in \cite{Ott2017}, this step is essential for the case that there are mechanical components other than the tracer moving periodically with time, as such kind of movements may lead to strong distortions to the signals from the tracer. Note that the smallest size of tracer that can be tracked is to a certain extent associated with the signal-to-noise, therefore special care has to be taken to remove the background noise for a better performance of the tracking system.

\begin{figure}
\centering
\includegraphics[width=0.95\columnwidth]{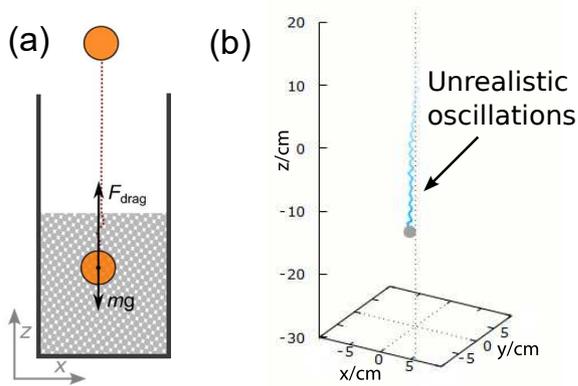}
\caption{An illustration (a) showing the trajectory of the tracer falling freely into a granular bed with a coordinate definition. (b) One segment of the reconstructed trajectory (corresponding to 0.04\,s). The unrealistic oscillations of the trajectories in both horizontal directions arise from IQ mismatch.}
\label{fig:osci}
\end{figure}

The third and perhaps the most serious source of error arises from the strong IQ mismatch, which is partly owing to the fact that the range of tracer movement (up to $0.5$\,m) is on the same order of magnitude as the working distance (about $1.5$\,m), consequently the fluctuations of both I and Q signals are relatively strong, leading to unrealistic fluctuations in the reconstructed trajectory. As an example, Figure\,\ref{fig:osci}(b) shows the reconstructed trajectory of a tracer falling freely under gravity. The oscillations in the horizontal direction can sometimes approach the diameter of the tracer. One possible reason in connection with the above description is the amplitude fluctuations in the IQ plane. The other possible reason is the the offset of IQ signals: The tracer movement corresponds to the phase shift in the IQ plane, which is determined by the corresponding tracer path, (ideally it should be an arc). For either case, there will be a systematic deviation of the measured phase angle with respect to the ideal one.

\begin{figure}
\centering
\includegraphics[width=0.95\columnwidth]{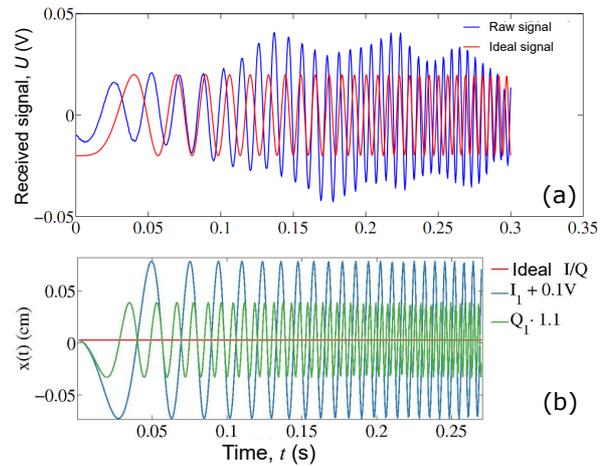}
\caption{(a) Real part of the raw signal from antenna 1 and one ideal case (without amplitude modulations) for comparison. (b) Reconstructed trajectory for the free falling case shown in \ref{fig:cali} for the ideal case (red line, without oscillations) and two artificially generated IQ mismatched cases: One with offset (blue curve) and the other one with Q signal being scaled up by a certain factor.}
\label{fig:mismatch}
\end{figure}

In order to demonstrate the above analysis, we introduce artificially the aforementioned two sources of error into the raw signal. As shown in Fig.\,\ref{fig:mismatch}(a), the raw signal of Q from the Rx. antenna 1 oscillates as a function of time, representing the fact that the tracer is moving away from the antennae when it falls down freely along the $z$ direction. The oscillation amplitude may fluctuates as time evolves, leading to one source of error. For the ideal case, one expects a constant amplitude (i.e., constant radius in the IQ plane), as indicated by the red curve. Using the ideal signal for reconstruction, one obtains a constant $x=0$, i.e., no unrealistic oscillations in the $x$ direction. If we artificially add a small offset to the I signal or multiple the Q signal with a factor slightly larger than 1 (i.e., introduce a slight distortion to the circular trajectory in the IQ plane). In either case, one observes clear oscillations in the $x$ direction, demonstrating how IQ mismatch leads to unrealistic fluctuations in the reconstructed trajectory. 

\begin{figure}
\centering
\includegraphics[width=0.8\columnwidth]{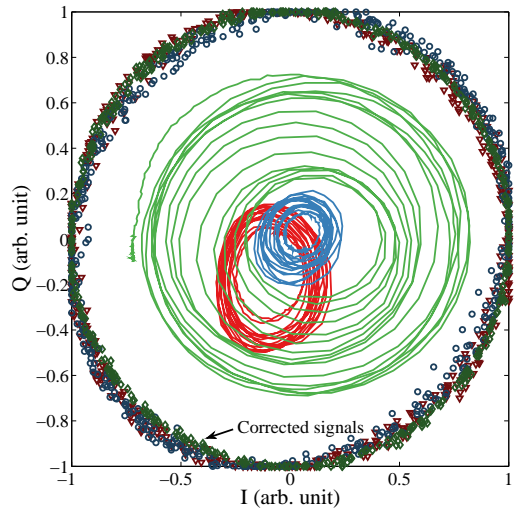}
\caption{Raw (continuous lines) and corrected (open symbols) signals representing a free-falling sphere from a height of $27$\,cm. Red (dark red), green (dark green) and blue (dark blue) curves (points) correspond to the results from channel 1, 2, and 3, respectively. For a better visibility, the offsets of the raw signals are removed. }
\label{fig:raw}
\end{figure}

\begin{figure}
\vskip 0em
\centering
\includegraphics[width=0.85\columnwidth]{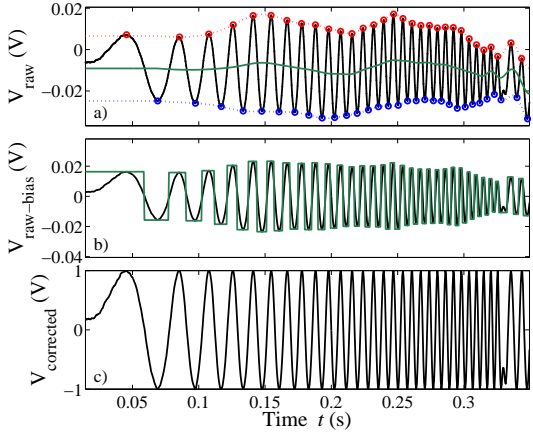}
\caption{Process for IQ mismatch correction. (a) A representative raw signal with peaks and valleys marked with red and blue open circles. From an average of both spline fits for the peaks and valleys, the bias error (green line) for the raw signal as a function of time is estimated. (b) Bias corrected signal time dependent rescaling factors (green line) for the correction of gain error. (c) Corrected signal for further analysis.}
\label{fig:corr}
\end{figure}

As Fig.\,\ref{fig:raw} shows, the raw IQ signals are typically not ideal in the sense that the IQ signals are not always orthogonal with each other. This mismatch may arise from the DC offsets of either I or Q signal, gain and phase imbalance. How to correct such kind of errors has been extensively discussed in, for instance \cite{Churchill1981} or \cite{Huang2002}, particularly along with the development of telecommunication and non-invasive motion detecting techniques \cite{Park2007}. The distortions are typically attributed to device imperfections as well as clutter. However, for the system being used here, there are additional errors arising from the mobility of the tracer itself, which can not be readily corrected with an additional calibration of the hardware. Moreover, distortion may also arise from the interaction of the scattered signal from the tracer with that from objects not completely transparent to EM waves. In that case, the existence of `mirrored' particles may lead to additional uncertainty.    

Here, we use the following approach to correct IQ mismatch arising from multiple sources of errors. It works best when the object moves in a distance covering multiple wavelengths. As illustrated in Fig.\,\ref{fig:corr}, the correction process is composed of the following steps: First, we identify the time segment of the raw data $V_{\rm raw}$ that contains the movement of the tracer particle via finding the start and end of the fluctuations. Second, the peaks (red circles) and valleys (blue circles) of individual fluctuations are determined by finding the local extreme values in the selected data. Third, the bias error $V_{\rm bias}$ [green line in (a)] is estimated as the mean value of the spline fits of peaks and valleys (dashed lines). In order to avoid unrealistic extrapolations, the bias error starts to vary only from the first peak. Fourth, the bias error is removed and the corrected signal $V_{\rm raw - bias}$ is segmented by zero crossings. Finally, the data in individual segments are rescaled by local maxim and minim to correct gain mismatch.   

As shown in Fig.\,\ref{fig:raw}, this approach can effectively find time dependent correction factors due to tracer movement. For the corrected data of channel 3 (dark blue circles), there exists a slight deviation from a perfect circle, indicating the existence of a small phase error. This arises presumably from the fact that perfect polarization cannot be achieved for all three Rx. antennae. Further investigations are needed to check whether this error can be avoided by using circular polarized EM waves or by correcting the phase error between I and Q signals in the Matlab program.

\begin{figure}
\vskip -2em
\centering
\includegraphics[width=0.85\columnwidth]{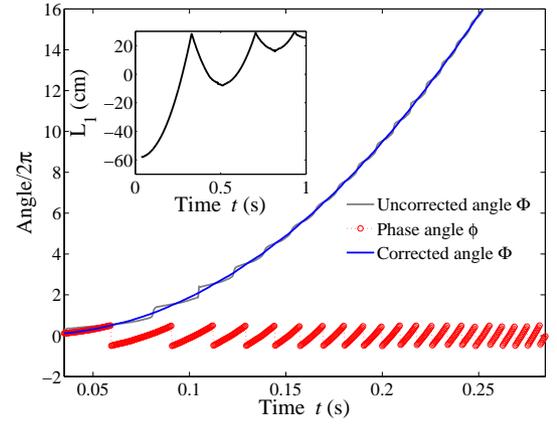}
\caption{Arc-tangential demodulation process to obtain the traveling distance from the Tx. to a Rx. antenna. The red open symbols correspond to the outcome from the demodulation and the blue curve represents the continuous phase shift that scales with the traveling distance of an EM wave. The gray curve corresponds to the $\Phi$ without correcting IQ mismatch. Inset shows an example of the variation of $L_{\rm 1}$ over a longer time.}
\label{fig:dist}
\end{figure}

After the correction of IQ mismatch, the corresponding phase angles are obtained by $\phi = \arctan(Q/I)$. Because $\phi$ is a modulo of $2\pi$, a further correction on the phase jump is needed to obtain the continuous phase shift $\Phi$. In this step, a threshold is introduced to determine whether a phase jump occurs or not and in which direction the jump takes place. As the phase shift of the $i$th channel $\Phi_i \propto L_i$, the variation of $\Phi$ with time (see the blue curve in Fig.\,\ref{fig:dist}) indicates that the target object moves initially slow and accelerates while moving away from the antennae. As demonstrated by a comparison between corrected $\Phi$ and uncorrected $\Phi_{\rm uncorr}$ phase shift, the aforementioned correction method can effectively reduce unrealistic fluctuations (see also Fig.\ref{fig:osci}(b)) in the reconstructed curves. More quantitatively, the magnitude of oscillations enhances with the speed of the projectile and it can reach $\pm 1.5$\,cm. After correction, it reduces to less than $1.0$\,mm. As shown in the inset of Fig.\,\ref{fig:dist}, the distance $L_1$ obtained from Rx. antenna represents exactly what is expected: The object falls freely with a growing velocity and bounces back when reaching the container bottom, suggesting that the coefficient of restitution, which measures the relative rebound over impact velocities, can be determined with the radar system. In comparison to the standard high speed imaging technique \cite{Mueller2016}, the radar tracking technique requires less data collection and processing efforts.

\section{Validation of correction algorithm}

\begin{figure}
\vskip -3em
\centering
\includegraphics[width=0.80\columnwidth]{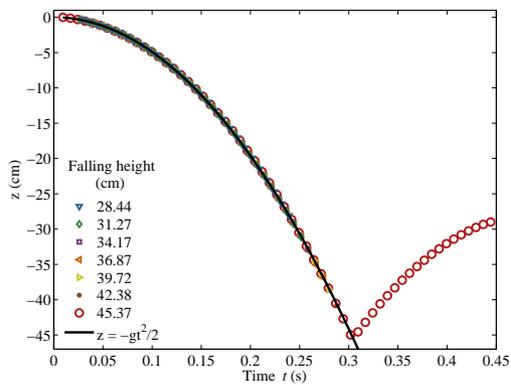}
\caption{A comparison of reconstructed free-falling curves at various initial falling heights. The solid line corresponds to the expected free-falling curve for the largest falling height. Note that the curves for various $H$ are shifted to have the initial falling position $z=0$\,cm, and only the trajectories before the first bouncing with the container bottom are shown except for $H=22.37$\,cm. For each curve, one over 15 data points are shown here for a better visibility. The typical error bar ($\sim 1$\,mm) of the position data is smaller than the symbol size.}
\label{fig:varHeight}
\end{figure}

As the goal of this investigation is to explore the possibility of obtaining an object moving inside a granular medium using microwave radar. We proceed with the following two steps:

First, we focus on a spherical projectile falling in free space without the presence of granular materials. We compare the reconstructed trajectories of the object from different initial falling heights with the expected parabolic curve. As shown in Fig.\,\ref{fig:varHeight}, the falling curves agree with the expected curve well, demonstrating that, after a proper correction of IQ mismatch, the radar system can be used for particle tracking. In particular, the tracer falls under gravity over a distance up to $50$\,cm, which is about $1/3$ of the working distance. After a proper correction of the signal fluctuations for both IQ channels, the correct information on the phase (i.e., distance) can be extracted to a satisfactory level. This outcome also suggests that the configuration built for releasing the tracer with minimized initial velocity and also for a better signal-to-noise ratio works well.

\begin{figure}
\centering
\includegraphics[width=\columnwidth]{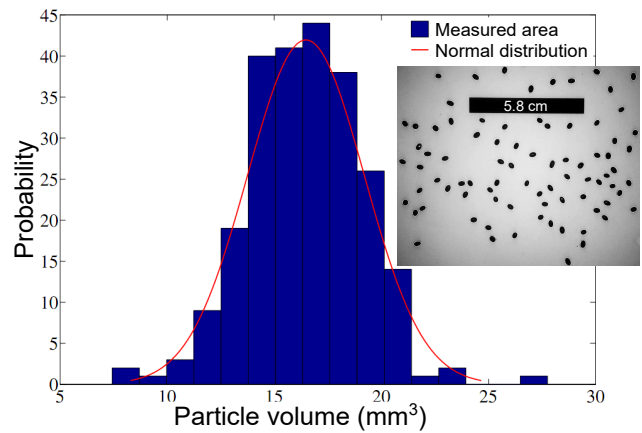}
\caption{A close view of expanded polypropylene (EPP) particles used as granular sample with the volume distribution assuming the particles are prolate ellipsoids. In total the shapes of 241 particles are analyzed to generate the distribution.}
\label{fig:granulates}
\end{figure}

Second, we replace the lower part of the free space with a granular medium composed of the EPP particles (see Fig.\,ref{setup}), which are expected to be transparent to EM waves. Expanded polypropylene (EPP) particles (Neopolen, P9255) are used as the granular sample. As the particles are porous with $95$\% air trapped inside, its dielectric constant (relative to vacuum) is $1.03$, very close to air. Therefore, they are practically transparent to electromagnetic (EM) waves. As the snapshot in Fig.\ref{fig:granulates} shows, the EPP particles have an ellipsoidal shape with a length-to-width ratio of $\approx 1.4$ and an average volume of $16.5\pm 2.7$\,mm, where the uncertainty corresponds to the standard deviation of the volume distribution. Here, the volume is estimated assuming that the particles are of prolate spheroid shape (i.e., $a=b<c$ with $a$, $b$ and $c$ semi-axes of an ellipsoid). The density of the particles is $92$\,kg $\cdot$ m$^{-3}$, which can typically be tuned in the expansion process. Other material properties include: average particle weight $1.2$\,mg, bulk density $55$\,kg/m$^3$, tensile strength $880$\,kPa, thermal conductivity $0.035$\,W/(mK).

\begin{figure}
\vskip 0em
\centering
\includegraphics[width=0.95\columnwidth]{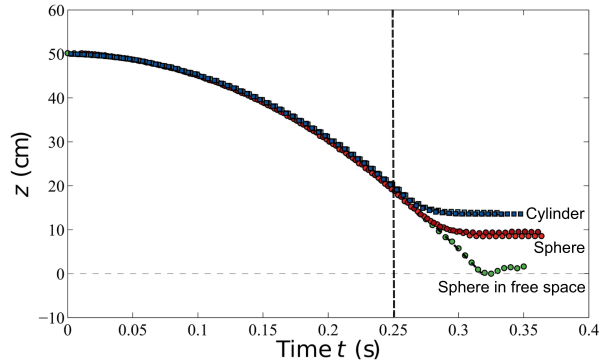}
\caption{Reconstructed trajectories of projectile impact into a granular bed composed of EPP particles. The initial falling height is fixed at $50$ cm with respect to the floor. The green, red and blue data points represent the three scenarios: Free falling into the empty container without granular filling, impact into a granular bed by a spherical tracer and by a cylindrical tracer particle. The typical error bar ($\sim 1$\,mm) of the position data is smaller than the symbol size. The vertical dashed line corresponds to the time when a projectile touches the surface of the granular medium.}
\label{fig:trj}
\end{figure}

Two intruder shapes are used for the test experiment: A sphere with a diameter of $1.45$\,cm and a cylinder with a diameter $1.55$\,cm and height $3.10$\,cm, both made of Styrofoam with the tracer embedded. For each type of projectile, three free-falling experiments are conducted with the same initial falling heights. As shown in Fig.\,\ref{fig:trj}, both projectiles follow the same trajectory before they touch the surface of the granular medium as expected. While penetrating through, cylindrical projectile experiences a slightly larger granular drag and consequently land at a relatively shallow depth in comparison to the spherical one. Note that the smoothness of the tracer trajectory while impacting on the granular layer demonstrates that the EPP particles chosen here are indeed transparent to EM waves and there is no influence from possible multiple scattering of EM waves by granular particles. For the case of spherical projectile, the reconstructed trajectories of ten repetitions give rise to the same parabola for the free falling region. This outcome validates the correction protocol introduced here. From the projectile trajectory inside a granular medium, one can obtain the acceleration as well as total force acting on it. Note that mechanical properties of granular particles can be determined experimentally or from the specifications provided by the producers. Subsequently those information can be used in numerical simulations using discrete element methods to have a direct comparison with experimental results. Thus, radar tracking technique can be used here to explore the `microscopic' origin of the drag force induced by granular particles in combination with numerical simulations. Interested readers may refer to \cite{Huang2020} for a recent example.

\section{CONCLUSION}

To summarize, this investigation suggests that advances in radar tracking technology can be helpful in the investigation of granular dynamics. Using an X-band continuous wave radar system, we are able to track a centimeter sized metallic object in 3D, which enables, for instance, a measurement of the coefficient of restitution of the particle. In comparison to other particle imaging techniques already being used for granular particles \cite{Amon2017}, continuous-wave radar tracking has the advantage of high time resolution and low data collection and processing requirements. With the rapid development of radar technology, this approach is also expected to be more cost effective and accurate. 

Moreover, we show that the accuracy of the radar tracking technique depends strongly on a proper correction of  IQ mismatch, which arises predominately from the mobility of the tracer itself. A practical approach has been proposed to correct the instantaneously changing bias as well as gain errors in the raw IQ signals. Finally, we validate this approach through an analysis on the reconstructed trajectories of projectiles falling under gravity in free space as well as impacting into a light granular medium. In comparison to our previous investigation\,\cite{Ott2017}, this approach enables more quantitative studies of an object moving in a three dimensional granular medium. 

Further investigations will focus on particle tracking with various types of granular materials, particularly how to deal with distorted signals arising from multiple scattering of the surrounding granular particles that are not completely transparent to EM waves. As the algorithm does not rely on the frequency band chosen, it would also be interesting to employ radar systems working at a higher frequency band to achieve better spatial resolution.

\section*{Acknowledgment}

We acknowledge Felix Ott for his preliminary work on the experimental set-up and Klaus Oetter for technical support. Helpful discussions with Valentin Dichtl, Simeon V\"olkel and Ingo Rehberg are gratefully acknowledged. This work is partly supported by German Research Foundation through Grant No.~HU1939/4-1.

\end{document}